\documentclass[twocolumn,aps,prl,floatfix,superscriptaddress]{revtex4-1}

\usepackage{amsmath,amssymb}
\usepackage{graphicx,float}
\usepackage{times,txfonts}
\usepackage{color}
\usepackage{multirow}
\usepackage{bbold}
\usepackage{color}
\usepackage{soul}
\usepackage{hyperref}
\usepackage{times,txfonts}
\usepackage{natbib}
\usepackage{nicefrac}
\usepackage{blindtext}
\usepackage[export]{adjustbox}
\usepackage{mathrsfs}
\usepackage{textcomp}
\usepackage{multirow}
\usepackage[normalem]{ulem}
\usepackage{amsmath}

\newcommand{\ket}[1]{\left| #1 \right\rangle}
\newcommand{\braket}[2]{\left\langle {#1{\left| \vphantom{#1 #2} \right.} #2} \right\rangle}
\newcommand{\qo}[1]{``#1''}

\renewcommand{\epsilon}{\varepsilon}

\def\VR{\kern-\arraycolsep\strut\vrule &\kern-\arraycolsep}
\def\vr{\kern-\arraycolsep & \kern-\arraycolsep}
\usepackage{soul}
\definecolor{lightblue}{RGB}{185,210,248}
\definecolor{lgreen}{RGB}{15,150,15}
\sethlcolor{lightblue}

\newcommand{\Fqb}{\textit{F}-qubit}
\newcommand{\Fqbs}{\textit{F}-qubits}

\begin{document}

\title{High-Dimensional Quantum Key Distribution with Qubit-like States}

\author{Lukas Scarfe}
\email{lscar039@uottawa.ca}
\affiliation{Nexus for Quantum Technologies, University of Ottawa, Ottawa ON, Canada, K1N 5N6}

\author{Rojan Abolhassani}
\affiliation{Nexus for Quantum Technologies, University of Ottawa, Ottawa ON, Canada, K1N 5N6}

\author{Frédéric Bouchard}
\affiliation{National Research Council of Canada, 100 Sussex Drive, Ottawa ON, Canada, K1A 0R6}

\author{Aaron Goldberg}
\affiliation{National Research Council of Canada, 100 Sussex Drive, Ottawa ON, Canada, K1A 0R6}

\author{Khabat Heshami}
\affiliation{Nexus for Quantum Technologies, University of Ottawa, Ottawa ON, Canada, K1N 5N6}
\affiliation{National Research Council of Canada, 100 Sussex Drive, Ottawa ON, Canada, K1A 0R6}

\author{Francesco Di Colandrea}
\affiliation{Nexus for Quantum Technologies, University of Ottawa, Ottawa ON, Canada, K1N 5N6}
\affiliation{Dipartimento di Fisica, Universit\`{a} degli Studi di Napoli Federico II, Complesso Universitario di Monte Sant'Angelo, Via Cintia, 80126 Napoli, Italy}

\author{Ebrahim Karimi}
\affiliation{Nexus for Quantum Technologies, University of Ottawa, Ottawa ON, Canada, K1N 5N6}
\affiliation{National Research Council of Canada, 100 Sussex Drive, Ottawa ON, Canada, K1A 0R6}
\affiliation{Institute for Quantum Studies, Chapman University, Orange, California 92866, USA}

\begin{abstract}
Quantum key distribution (QKD) protocols most often use two conjugate bases in order to verify the security of the quantum channel. In the majority of protocols, these bases are mutually unbiased to one another, which is to say they are formed from balanced superpositions of the entire set of states in the opposing basis. Here, we introduce a high-dimensional QKD protocol using qubit-like states, referred to as Fourier-qubits (or \Fqbs{}). In our scheme, each \Fqb{} is a superposition of only two computational basis states with a relative phase that can take $d$ distinct values, where $d$ is the dimension of the computational basis. This non-mutually unbiased approach allows us to bound the information leaked to an eavesdropper, maintaining security in high-dimensional quantum systems despite the states' seemingly two-dimensional nature. By simplifying state preparation and measurement, our protocol offers a practical alternative for secure high-dimensional quantum communications. We experimentally demonstrate this protocol for a noisy high-dimensional QKD channel using the orbital angular momentum degree of freedom of light and discuss the potential benefits for encoding in other degrees of freedom. 
\end{abstract}	

\maketitle

\section{Introduction} 
Quantum Key Distribution (QKD) promises an information-theoretically secure method to distribute shared keys guaranteed by the fundamental laws of physics~\cite{BB84,1176619,Zahidy2024}. 

In order to distribute a secure key over an untrusted channel, Alice and Bob must prepare, exchange, and measure quantum states with few errors, which are not trivial tasks~\cite{Zhang_2025}. This difficulty has spawned many QKD protocols that can be more easily implemented with realistic devices, and improve the secure key rates despite the practical challenges~\cite{Stucki2005,MDIQKD,Sasaki2014}.  

High-dimensional (HD) QKD protocols have attracted growing interest in recent years~\cite{PhysRevApplied.14.014051, Sit:17}. Indeed, by encoding information into a higher-dimensional Hilbert space, secure keys can be distributed with a higher density of information per photon~\cite{bechmann2000quantum}.
In addition, these protocols can tolerate a greater error rate introduced by either an eavesdropper or simple channel noise~\cite{PhysRevLett.88.127902}.
While these benefits are enticing, the reality is that, up to this point, nearly all commercial QKD systems operate with two-dimensional QKD protocols~\cite{6459842}. Likewise, most fundamental research is focused on two-dimensional implementations~\cite{Bedington2017}. The main reason is that encoding information in a high-dimensional Hilbert space poses significant technical challenges~\cite{HDQKDReview}, 
with the complexity of experimental setups scaling proportionally to the dimensionality of the protocol. In contrast, a two-dimensional QKD system can process the polarisation degree of freedom, offering the advantage of cost-effective and straightforward passive generation and detection equipment.

The go-to QKD protocol, BB-84, requires two mutually unbiased bases (MUB), where each basis consists of states in a balanced superposition of every state in the opposing basis. The two most common bases chosen are the computational (logical) basis and the Fourier basis~\cite{PhysRevApplied.11.064058,PhysRevLett.88.127902,10.1063/5.0185281}.
In the $d$-dimensional implementation, the use of the Fourier basis requires precisely generating and measuring superpositions of $d$ states. 

In this work, we introduce a set of \qo{qubit-like} states to be used in QKD alongside the logical basis in lieu of the Fourier basis.
These states are qubit-like in the sense that they are only constructed as superpositions of two logical states, with a relative phase difference between them that is one of the $d$ roots of unity. Because of the binary nature of these states, and their relation to the well-known quantum Fourier transform (QFT), we refer to these states as Fourier-qubits (or \Fqbs{}).
This reduction in the complexity of the states allows for simpler generation and detection~\cite{Wang_2018}.
While previous works with qubit-like states used them to increase the error tolerance up to the theoretical limit of $50\%$, the key rate remains that of a two-dimensional protocol, that is 1 bit per sifted photon~\cite{Chau15,Chau15Experiment}.
Here, the \Fqbs{} are used to complement the logical basis, whose generation and detection are the most simple, and estimate the information leaked to a potential eavesdropper. This feature allows our QKD protocol to maintain the high-dimensional benefit of increased information density of $\log_2(d)$ bits per sifted photon while using relatively simple quantum states. 

We theoretically prove that our protocol is secure from the most general coherent attack by an eavesdropper by demonstrating that the \Fqbs{}  can be used to indirectly measure the phase error rate. Additionally, we experimentally demonstrate the generation and detection of these modes in a noisy lab-scale channel using spatial modes of light in a 4-dimensional Hilbert space. In our test, the channel supports our \Fqb{}-based QKD protocol with a measured sifted key rate above $1$ bit per sifted photon. We foresee our protocol being useful in implementations where high-dimensional QKD is desirable such as high-bitrate and low-noise channels that are bandwidth limited by the detector recovery time. Additionally, any system where the complexity of the quantum state scales with dimensionality will find benefit from the implementation of the \Fqbs{}.

\section{Results}
\subsection{Protocol}
In our protocol, information is encoded onto qudits living in a $d$-dimensional Hilbert space. In particular, Alice and Bob prepare and measure their qudits in two separate bases. In the logical basis, information is encoded in states of the form ${\ket{\psi_n}=\ket{n}}$, where {$n \in \{0,1,...,d-1 \}$}. Alice randomly selects $n$, serving as her raw key, and transmits the state $\ket{n}$ over an untrusted channel. Upon reception, Bob projects its incoming state in the same basis and uses the measurement outcome ${n'}$ as his raw key. By iterating this process, Alice and Bob generate a raw key and use an authenticated classical channel to compare a small subset of their keys, estimating the error rate $E_d$ to perform error reconciliation.

\begin{figure}
    \centering
    \includegraphics[width=1.0\columnwidth]{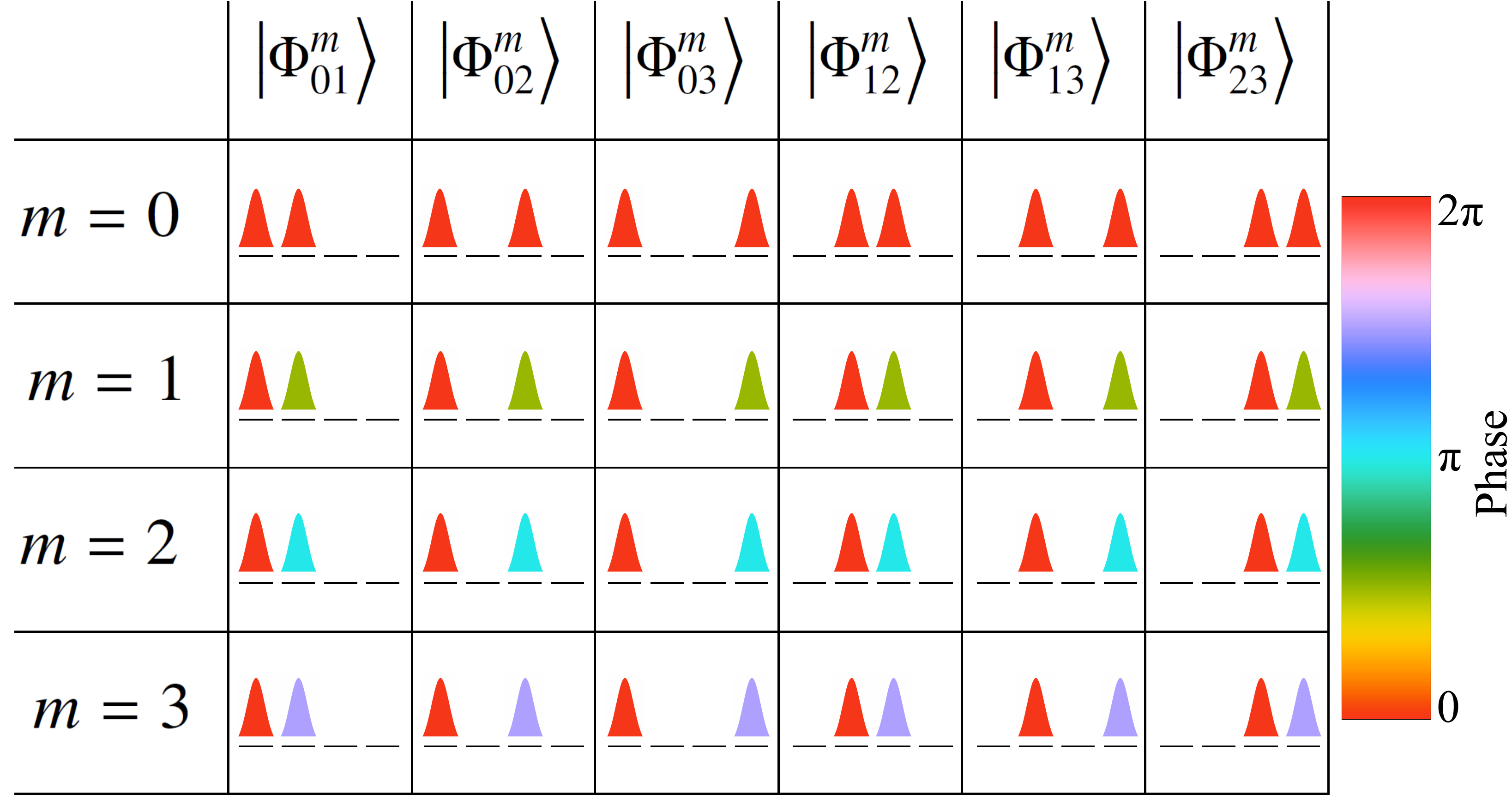}
		\caption{\textbf{\Fqbs{} in a 4-dimensional time-bin-based protocol.} The logical basis consists of chopped Gaussian modes that are separated into time-bins $\ket{0}$, $\ket{1}$, $\ket{2}$, $\ket{3}$. False colours encode the relative phase information.} 
		\label{fig:4Dpubstime}
\end{figure}

In order to bound Eve's leaked information, Alice and Bob also prepare and measure states in a second basis containing the \Fqb{} states:
\begin{equation}
    |\phi_{jk}^{m} \rangle = (|j\rangle + \omega_d^m|k\rangle)/\sqrt{2},
\end{equation}
where ${\omega_d= e^{2 \pi i /d }}$, ${j \in\{0,1...,d-2\}}$, ${k\in\{1,2,...,d-1\}}$, with ${\: j< k}$, and ${m\in\{0,1,...,d-1\}}$. A depiction of these modes in a time-bin implementation can be seen in Fig.~\ref{fig:4Dpubstime}. More precisely, Alice randomly selects a triplet ${(j,k,m)}$, preparing the state ${\ket{\phi_{jk}^{m}}}$. Bob performs a measurement in the same \Fqb{} basis, projecting onto $\ket{\phi_{j'k'}^{m'}}$. Via classical communications, Alice and Bob can determine the probability of obtaining errors in the \Fqb{} basis. This information is used to estimate the phase error rate, $E'_d$, to bound the information leaked to an eavesdropper as in BB84. 

Finally, as in other high-dimensional BB84-like protocols, the secret key rate per sifted photon is given by~\cite{Sheridan:2010}

\begin{equation}
    R =  \log_2(d) - h^{(d)} (E_d) - h^{(d)} (E'_d),
      \label{eqn:secretrate}
\end{equation}
where ${h^{(d)}(x)= -x\log_2(x/(d-1))-(1-x)\log_2(1-x)}$ is the $d$-dimensional Shannon entropy function, and $E_d$ and $E'_d$ are the channel dit error and phase error rates, respectively. 

\begin{figure}
    \centering
    \includegraphics[width=1.0\columnwidth]{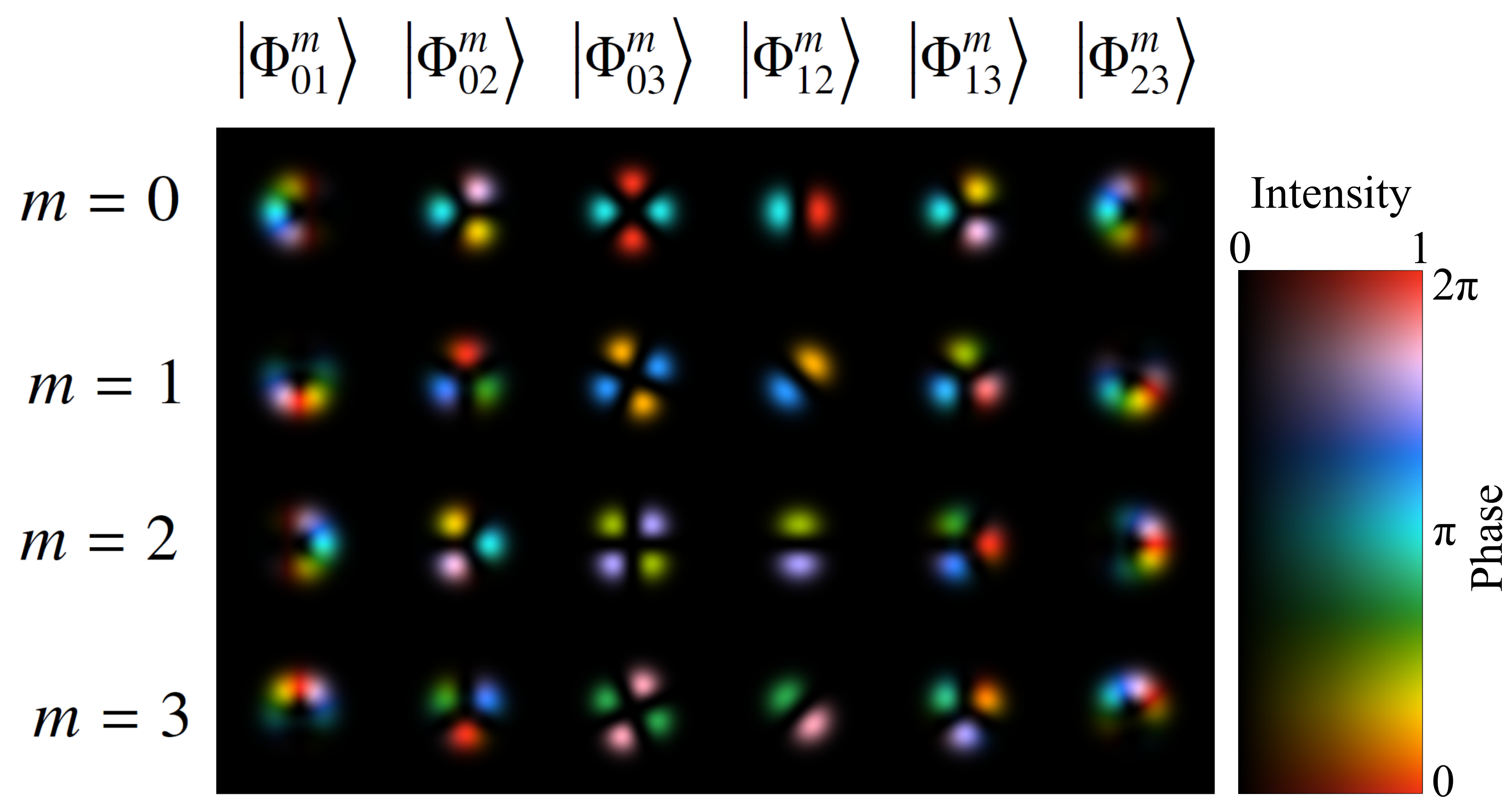}
		\caption{\textbf{\Fqbs{} in a 4-dimensional orbital angular momentum (OAM) based protocol.} The logical basis consists of a set of LG modes, carrying discrete units of OAM, $\ket{\ell}, \: \ell \in \{-2,-1,1,2\}$. False colours and opacity encode the phase and amplitude information, respectively.} 
		\label{fig:4Dpubs}
\end{figure}
\begin{figure*}[!htb]
	\begin{center}
		\includegraphics[width=1.2\columnwidth]{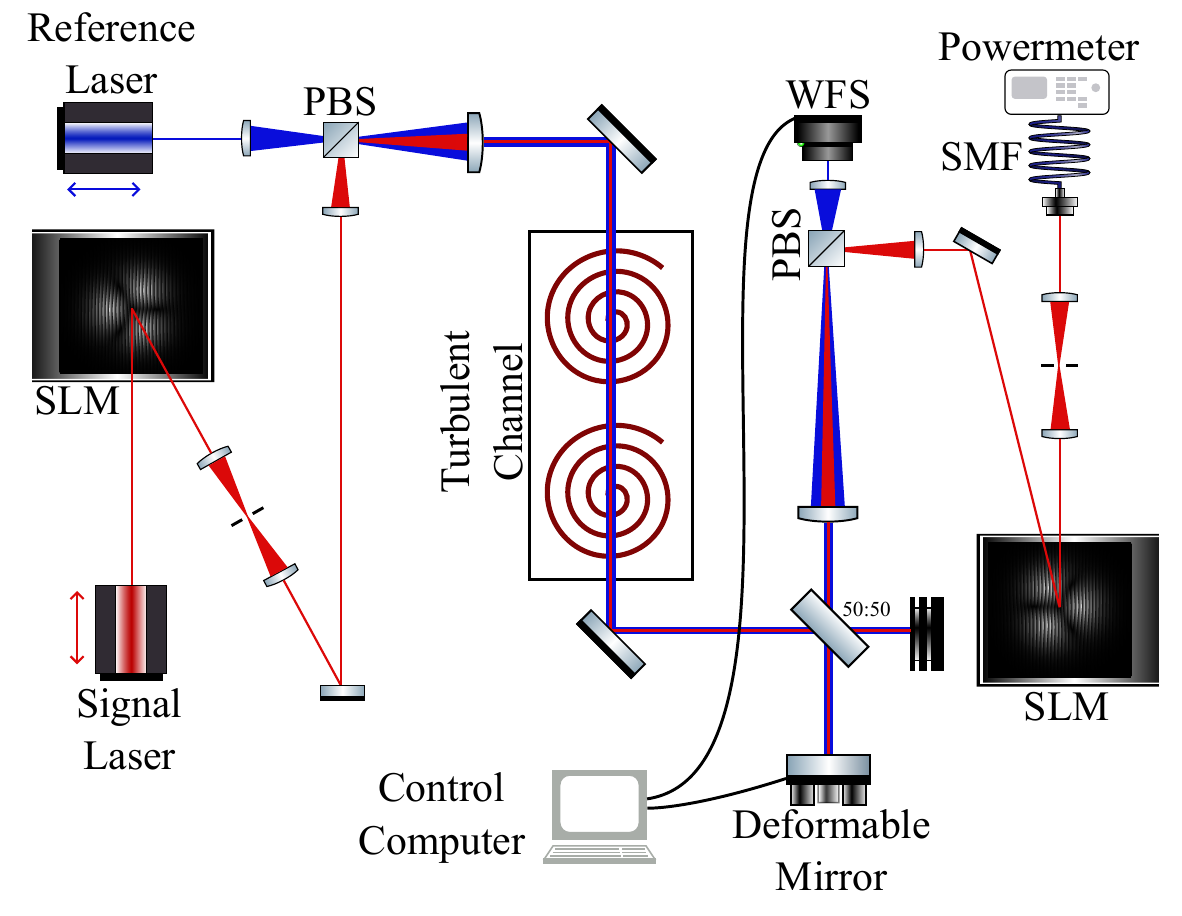}
		\caption{\textbf{Experimental setup to generate and measure the \Fqb{} modes encoded using OAM in a noisy channel.} The reference and signal laser are both of wavelength $633$ nm, with the red and blue representing orthogonal polarisations. The mode of the reference beam is a Gaussian, expanded to approximate a plane wave that is used as a probe of the turbulence measured by the wavefront sensor (WFS). The signal laser is impinged on a spatial light modulator (SLM) which is used to apply the phase and intensity of the \Fqb{} modes encoded as a superposition of LG modes carrying OAM. These beams are made to propagate co-linearly using a polarizing beam splitter (PBS) and sent through the turbulent channel. They are subsequently separated after being corrected by the adaptive optics mirror. The \Fqb{} mode is projectively measured using a second SLM and coupled into a single-mode fibre (SMF).}
		\label{fig:setup}
	\end{center}
\end{figure*}

\subsection{Proof of Security}\label{Proof}

Assuming the transmitted qudit is generated by an ideal single-photon source, the action of an eavesdropper, Eve, can be generally modeled as a coherent attack given by the unitary transformation $U_\mathrm{Eve}$:

\begin{equation}
    U_\mathrm{Eve} |\eta_k\rangle |e_{00}\rangle = \sum_{j=0}^{d-1} c_{k j} |\eta_j\rangle |e_{k j}\rangle,
\end{equation}
where ${\ket{e_{k j}}}$ is Eve's ancilla state and ${\ket{\eta_k}}$ is extracted from the Fourier-conjugate basis:
\begin{equation}\label{eq:fourierbasis}
    {|\eta_k \rangle=\frac{1}{\sqrt{d}}\sum_{n=0}^{d-1} \omega_d^{k n} |n\rangle }.
\end{equation}
Without loss of generality, we assume ${c_{kj} \geq 0}$ and ${\langle e_{kj}|e_{rs}\rangle=\delta_{kr}\delta_{js}}$. 

The dit error rate of the Fourier basis, $E_d'$, depends on the probability that Bob obtains a measurement of ${|\eta_{n} \rangle}$ conditioned on the fact that Alice prepares ${|\eta_{\ell} \rangle}$:

\begin{eqnarray}
   && p(\eta_{n} |\eta_{\ell}) \\
   \nonumber && = \mathrm{Tr} \left[ |\eta_{n}\rangle \langle \eta_{n}| U_\mathrm{Eve} |\eta_{\ell}\rangle \langle \eta_{\ell}| \otimes |e_{00}\rangle \langle e_{00}| U_\mathrm{Eve}^\dagger \right] \\
   \nonumber && = c_{\ell n}^2.
\end{eqnarray}

We are now interested in relating the probability outcomes $p(\eta_{n}|\eta_{\ell})$ in the Fourier basis to the measured probability outcomes of the qubit-like states. To do so, we use the fact that the \Fqb{} states can be rewritten in terms of the Fourier basis states as
\begin{equation}
    |\phi_{jk}^{m}\rangle = \frac{1}{\sqrt{2d}}\sum_\ell \left( \omega_d^{-j \ell}+\omega_d^{(m-k \ell)} \right) |\eta_\ell \rangle.
\end{equation}
The probability that Bob obtains $|\phi_{j'k'}^{m'}\rangle$ conditioned on the fact that Alice prepares ${|\phi_{jk}^{m}\rangle}$ is thus given by

\begin{eqnarray} \label{eq:coeffsfour-Fqb}
    && p(\phi_{j'k'}^{m'}|\phi_{jk}^{m}) \\ \nonumber
     \nonumber && = \frac{4}{d^2} \sum_{\ell n} \cos^2 \left[\frac{\pi(m-(k-j)\ell)}{d} \right] \\ && \nonumber \ \ \ \ \ \ \ \ \ \ \ \ \ \ \  \cos^2 \left[\frac{\pi(m'-(k'-j')n)}{d} \right]  p(\eta_n|\eta_\ell).\\ \nonumber
\end{eqnarray}

This relation can be inverted to find the probability outcomes $p(\eta_{n}|\eta_{\ell})$ in the Fourier basis given the probability outcomes of the \Fqbs{},

\begin{eqnarray} \label{eq:coeffs}
   && p(\eta_n|\eta_\ell) = \frac{4}{d^4} \sum_{j<k}\sum_{j'<k'}\sum_{m,m'} \left( \beta + \sum_{p=0}^{d-1} \exp \left( \frac{2 \pi i}{d} p \left( m - (k-j)\ell \right) \right) \right) \nonumber \\
    && \left( \beta + \sum_{p'=0}^{d-1} \exp \left( \frac{2 \pi i}{d} p' \left( m' - (k'-j')n \right) \right) \right) p(\phi_{j'k'}^{m'}|\phi_{jk}^{m}),
\end{eqnarray} 
where $\beta=(2-d)/(d-1)$. For the full derivation of this relation, refer to the accompanying supplementary material.
Recall that the result of Eq.~\eqref{eqn:secretrate} is expressed in terms of dit error rate ($E_d$) and the phase error rate ($E'_d$) of the computational basis. 
Eve's leaked information 
is given by $I_\mathrm{AE}\leq h^{(d)}(E'_d)$~\cite{shor2000simple}, where ${I_\text{AE}}$ is the mutual information between Alice and Eve. Explicitly, the dit error rate in the Fourier basis is given by,

\begin{eqnarray} \label{eq:Phaseerr}
    E_d' &=& \frac{1}{d} \sum_{\ell \neq n} p(\eta_n|\eta_\ell).
\end{eqnarray}

Using the fact that the dit error rate in the Fourier basis is equal to the phase error in the computational basis~\cite{lo2005efficient}, we have then linked the probability of detection matrices of the \Fqbs{}  to phase error rate of the computational basis. With both the dit and phase error rates in the channel, Alice and Bob can perform error correction and generate a secure key so long as both errors are below the tolerable threshold~\cite{PhysRevLett.85.441}. 

Finally, the secret key rate per sifted photon is given by
\begin{eqnarray}
    R &=& \log_2(d) - h^{(d)} (E_d) - I_\mathrm{AE} \\ \nonumber & \geq & \log_2(d) - h^{(d)} (E_d) - h^{(d)} (E'_d).
\end{eqnarray}

\subsection{Implementation}
We test the generation and detection of these modes in a hypothetical orbital angular momentum (OAM) based protocol through a noisy short-distance free-space optical channel. In particular, we use Laguerre-Gaussian (LG) beams, each carrying a discrete value of OAM, $\ell$~\cite{PhysRevA.45.8185}, forming a 4-dimensional Hilbert space. The phase and intensity of the spatial modes are shown and labeled in Fig.~\ref{fig:4Dpubs}. The \Fqb{} modes are generated by impinging a $633$ nm beam from a HeNe laser onto a spatial light modulator (SLM) displaying a grated hologram designed to precisely control phase and intensity~\cite{Bolduc:13} of an output beam. This beam along with a Gaussian beam of the same wavelength and opposite polarisation are sent through a channel with noise simulated by a turbulent cell, as well as an adaptive optics system as a corrective element~\cite{Scarfe2025}. A simplified experimental setup is shown in Fig.~\ref{fig:setup}. Further information on the use of the noisy channel and the adaptive optics system, can be found in the included supplementary material.

\begin{figure}
    \centering
    \includegraphics[width=1.0\columnwidth]{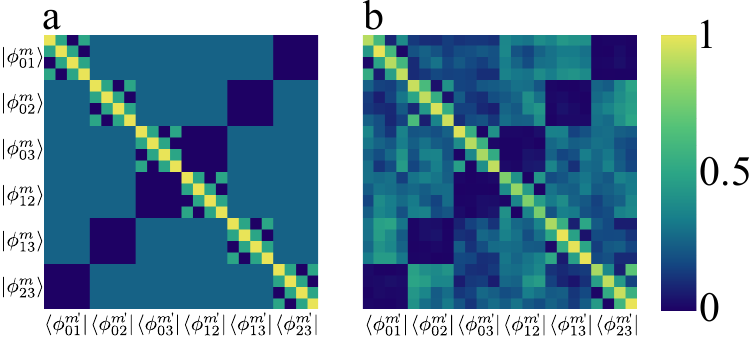}
		\caption{\textbf{Probability of detection matrices of \Fqb{} basis in 4 dimensions with OAM encoding.} Each subgroup of $4\times4$ is representative of $\Big|\braket{\phi_{i,j}^m}{\phi_{i',j'}^{m'}}\Big|^2$, with $m,m'$ varying from $0$ to $d-1$. \textbf{a)} Theoretical probability of detection matrix with no errors. \textbf{b)} Measured probability of detection matrix through our noisy channel.
  }\label{fig:crosstalks}
\end{figure}

The \Fqb{} mode is sent to another SLM where a projective measurement is performed using intensity and phase masking~\cite{Bouchard:18}. Each generated mode is projectively measured on all states. These measurements are normalized to create a probability of detection matrix shown in Fig.~\ref{fig:crosstalks} according to the procedure outlined in the accompanying supplementary material. This probability of detection matrix of the \Fqb{} basis is then used to calculate the phase error rate as shown in Eqs.~\eqref{eq:coeffs},~\eqref{eq:Phaseerr}. We calculate the phase error rate, $E_d'$, by minimizing the distance between left and right sides of Eq.~\eqref{eq:coeffsfour-Fqb}. This minimization is performed in order to avoid unphysical solutions present due to detector noise. Using measurements of the computational basis, we directly determine the bit error rate in the channel.

In our noisy channel, we find that the dit error rate, $E_d$, is $2.89\%$, and the phase error rate, $E'_d$, is $6.91\%$. After error correction and privacy amplification, the resulting sifted key rate in this channel as given in Eq.~\eqref{eqn:secretrate}, would be $R=1.28$ bits per photon.

\section{Conclusions}
We have introduced a high-dimensional quantum key distribution protocol that uses qubit-like states in the secondary basis. Due to their qubit-like nature, the preparation and detection of \Fqbs{} is significantly less complex than traditional MUB states, while maintaining the benefits of high-dimensional protocols, namely the increase in information density and error tolerance. We have experimentally implemented these qubit-like modes in a noisy lab-scale OAM-based QKD channel achieving a sifted key rate above $1$ bit per sifted photon. 

The \Fqb{} basis is overcomplete and is constructed of $d^2(d-1)/2$ states, with a factor of $d^2$ more states than the traditional BB-84 protocol. This is not a problem in the limit of infinite key length, but it should be considered for real-world implementations. Nevertheless, this issue can be alleviated with an unbalanced basis choice~\cite{lo2005efficient}, wherein the \Fqbs{} are only used to estimate the error rate of the channel and not for key generation and are generated at a lower rate than the logical basis.

A common issue facing QKD systems is that the bandwidth of the channel is restricted by the limitation of single photon detector recovery times~\cite{10.1126/sciadv.1701491}. These types of channels are particularly suited to implementing high dimensional QKD in order to increase the information density per photon, meanwhile the simplicity of the \Fqbs{} makes them useful in many different high-dimensional encoding schemes. For example, the implementation in a time-bin encoding simply requires a variable length interferometer. We feel that these modes will have use in fields other than quantum communications, such as quantum sensing and imaging. We will continue to further investigate potential applications.

\vspace{0.5cm}
\noindent \textbf{Acknowledgments.}
This work was supported by Canada Research Chairs; Canada First Research Excellence Fund (CFREF); National Research Council of Canada High-Throughput and Secure Networks (HTSN) Challenge Program; and the Alliance Consortia Quantum Grant (QUINT, ARAQNE). Francesco Di Colandrea further acknowledges support from the PNRR MUR project PE0000023-NQSTI.\\

\noindent \textbf{Disclosures} The authors declare no conflicts of interest.\\

\noindent \textbf{Data availability} Data underlying the results presented in this paper may be obtained from the authors upon reasonable request.


\newpage
\bibliographystyle{ieeetr}
\bibliography{FQubit}

\providecommand{\noopsort}[1]{}
\begin{thebibliography}{10}

\bibitem{BB84}
C.~H. Bennett and G.~Brassard, ``Quantum cryptography: Public key distribution and coin tossing,'' {\em Theoretical Computer Science}, vol.~560, pp.~7--11, 2014.
\newblock Theoretical Aspects of Quantum Cryptography – celebrating 30 years of BB84.

\bibitem{1176619}
D.~Gottesman and H.-K. Lo, ``Proof of security of quantum key distribution with two-way classical communications,'' {\em IEEE Transactions on Information Theory}, vol.~49, no.~2, pp.~457--475, 2003.

\bibitem{Zahidy2024}
M.~Zahidy, D.~Ribezzo, C.~De~Lazzari, I.~Vagniluca, N.~Biagi, R.~M{\"u}ller, T.~Occhipinti, L.~K. Oxenl{\o}we, M.~Galili, T.~Hayashi, D.~Cassioli, A.~Mecozzi, C.~Antonelli, A.~Zavatta, and D.~Bacco, ``Practical high-dimensional quantum key distribution protocol over deployed multicore fiber,'' {\em Nature Communications}, vol.~15, p.~1651, Feb 2024.

\bibitem{Zhang_2025}
H.~Zhang, W.~Li, R.~He, Y.~Zhang, F.~Xu, and W.~Gao, ``Noise-reducing quantum key distribution,'' {\em Reports on Progress in Physics}, vol.~88, p.~016001, dec 2024.

\bibitem{Stucki2005}
D.~Stucki, N.~Brunner, N.~Gisin, V.~Scarani, and H.~Zbinden, ``{Fast and simple one-way quantum key distribution},'' {\em Applied Physics Letters}, vol.~87, p.~194108, 11 2005.

\bibitem{MDIQKD}
Z.-Q. Yin, C.-H.~F. Fung, X.~Ma, C.-M. Zhang, H.-W. Li, W.~Chen, S.~Wang, G.-C. Guo, and Z.-F. Han, ``Measurement-device-independent quantum key distribution with uncharacterized qubit sources,'' {\em Physical Review A}, vol.~88, p.~062322, Dec 2013.

\bibitem{Sasaki2014}
T.~Sasaki, Y.~Yamamoto, and M.~Koashi, ``Practical quantum key distribution protocol without monitoring signal disturbance,'' {\em Nature}, vol.~509, pp.~475--478, May 2014.

\bibitem{PhysRevApplied.14.014051}
I.~Vagniluca, B.~Da~Lio, D.~Rusca, D.~Cozzolino, Y.~Ding, H.~Zbinden, A.~Zavatta, L.~K. Oxenl\o{}we, and D.~Bacco, ``Efficient time-bin encoding for practical high-dimensional quantum key distribution,'' {\em Physical Review Applied}, vol.~14, p.~014051, Jul 2020.

\bibitem{Sit:17}
A.~Sit, F.~Bouchard, R.~Fickler, J.~Gagnon-Bischoff, H.~Larocque, K.~Heshami, D.~Elser, C.~Peuntinger, K.~G\"{u}nthner, B.~Heim, C.~Marquardt, G.~Leuchs, R.~W. Boyd, and E.~Karimi, ``High-dimensional intracity quantum cryptography with structured photons,'' {\em Optica}, vol.~4, pp.~1006--1010, Sep 2017.

\bibitem{bechmann2000quantum}
H.~Bechmann-Pasquinucci and W.~Tittel, ``Quantum cryptography using larger alphabets,'' {\em Physical Review A}, vol.~61, no.~6, p.~062308, 2000.

\bibitem{PhysRevLett.88.127902}
N.~J. Cerf, M.~Bourennane, A.~Karlsson, and N.~Gisin, ``Security of quantum key distribution using $\mathit{d}$-level systems,'' {\em Physical Review Letters}, vol.~88, p.~127902, Mar 2002.

\bibitem{6459842}
L.~Oesterling, D.~Hayford, and G.~Friend, ``Comparison of commercial and next generation quantum key distribution: Technologies for secure communication of information,'' in {\em 2012 IEEE Conference on Technologies for Homeland Security (HST)}, pp.~156--161, 2012.

\bibitem{Bedington2017}
R.~Bedington, J.~M. Arrazola, and A.~Ling, ``Progress in satellite quantum key distribution,'' {\em npj Quantum Information}, vol.~3, p.~30, Aug 2017.

\bibitem{HDQKDReview}
D.~Cozzolino, B.~Da~Lio, D.~Bacco, and L.~K. Oxenløwe, ``High-dimensional quantum communication: Benefits, progress, and future challenges,'' {\em Advanced Quantum Technologies}, vol.~2, no.~12, p.~1900038, 2019.

\bibitem{PhysRevApplied.11.064058}
D.~Cozzolino, D.~Bacco, B.~Da~Lio, K.~Ingerslev, Y.~Ding, K.~Dalgaard, P.~Kristensen, M.~Galili, K.~Rottwitt, S.~Ramachandran, and L.~K. Oxenl\o{}we, ``Orbital angular momentum states enabling fiber-based high-dimensional quantum communication,'' {\em Physical Review Applied}, vol.~11, p.~064058, Jun 2019.

\bibitem{10.1063/5.0185281}
A.~Forbes, M.~Youssef, S.~Singh, I.~Nape, and B.~Ung, ``Quantum cryptography with structured photons,'' {\em Applied Physics Letters}, vol.~124, p.~110501, 03 2024.

\bibitem{Wang_2018}
S.~Wang, Z.-Q. Yin, H.~F. Chau, W.~Chen, C.~Wang, G.-C. Guo, and Z.-F. Han, ``Proof-of-principle experimental realization of a qubit-like qudit-based quantum key distribution scheme,'' {\em Quantum Science and Technology}, vol.~3, p.~025006, mar 2018.

\bibitem{Chau15}
H.~F. Chau, ``Quantum key distribution using qudits that each encode one bit of raw key,'' {\em Physical Review A}, vol.~92, p.~062324, Dec 2015.

\bibitem{Chau15Experiment}
H.~F. Chau, Q.~Wang, and C.~Wong, ``Experimentally feasible quantum-key-distribution scheme using qubit-like qudits and its comparison with existing qubit- and qudit-based protocols,'' {\em Physical Review A}, vol.~95, p.~022311, Feb 2017.

\bibitem{Sheridan:2010}
L.~Sheridan and V.~Scarani, ``{Security proof for quantum key distribution using qudit systems},'' {\em Physical Review A}, vol.~82, no.~3, pp.~1--4, 2010.

\bibitem{shor2000simple}
P.~W. Shor and J.~Preskill, ``Simple proof of security of the bb84 quantum key distribution protocol,'' {\em Physical Review Letters}, vol.~85, no.~2, p.~441, 2000.

\bibitem{lo2005efficient}
H.-K. Lo, H.~F. Chau, and M.~Ardehali, ``Efficient quantum key distribution scheme and a proof of its unconditional security,'' {\em Journal of Cryptology}, vol.~18, pp.~133--165, Apr 2005.

\bibitem{PhysRevLett.85.441}
P.~W. Shor and J.~Preskill, ``Simple proof of security of the bb84 quantum key distribution protocol,'' {\em Physical Review Letters}, vol.~85, pp.~441--444, Jul 2000.

\bibitem{PhysRevA.45.8185}
L.~Allen, M.~W. Beijersbergen, R.~J.~C. Spreeuw, and J.~P. Woerdman, ``Orbital angular momentum of light and the transformation of laguerre-gaussian laser modes,'' {\em Physical Review A}, vol.~45, pp.~8185--8189, Jun 1992.

\bibitem{Bolduc:13}
E.~Bolduc, N.~Bent, E.~Santamato, E.~Karimi, and R.~W. Boyd, ``Exact solution to simultaneous intensity and phase encryption with a single phase-only hologram,'' {\em Optics Letters}, vol.~38, pp.~3546--3549, Sep 2013.

\bibitem{Scarfe2025}
L.~Scarfe, F.~Hufnagel, M.~F. Ferrer-Garcia, A.~D'Errico, K.~Heshami, and E.~Karimi, ``Fast adaptive optics for high-dimensional quantum communications in turbulent channels,'' {\em Communications Physics}, vol.~8, p.~79, Feb 2025.

\bibitem{Bouchard:18}
F.~Bouchard, N.~H. Valencia, F.~Brandt, R.~Fickler, M.~Huber, and M.~Malik, ``Measuring azimuthal and radial modes of photons,'' {\em Optics Express}, vol.~26, pp.~31925--31941, Nov 2018.

\bibitem{10.1126/sciadv.1701491}
N.~T. Islam, C.~C.~W. Lim, C.~Cahall, J.~Kim, and D.~J. Gauthier, ``Provably secure and high-rate quantum key distribution with time-bin qudits,'' {\em Science Advances}, vol.~3, no.~11, p.~e1701491, 2017.

\end{thebibliography}

\clearpage
\onecolumngrid
\renewcommand{\figurename}{\textbf{Figure}}
\setcounter{figure}{0} \renewcommand{\thefigure}{\textbf{S{\arabic{figure}}}}
\setcounter{table}{0} \renewcommand{\thetable}{S\arabic{table}}
\setcounter{section}{0} \renewcommand{\thesection}{S\arabic{section}}
\renewcommand{\thesubsection}{S.\arabic{subsection}}
\setcounter{equation}{0} \renewcommand{\theequation}{S\arabic{equation}}
\onecolumngrid
\vspace{1 EM}

\section{Supplementary Information for: ``High-Dimensional Quantum Key Distribution with Qubit-like States"}

\subsection{Inspiration}\label{Supp:insp}
\begin{figure}[htb!]
    \centering
    \includegraphics[width=0.60\columnwidth]{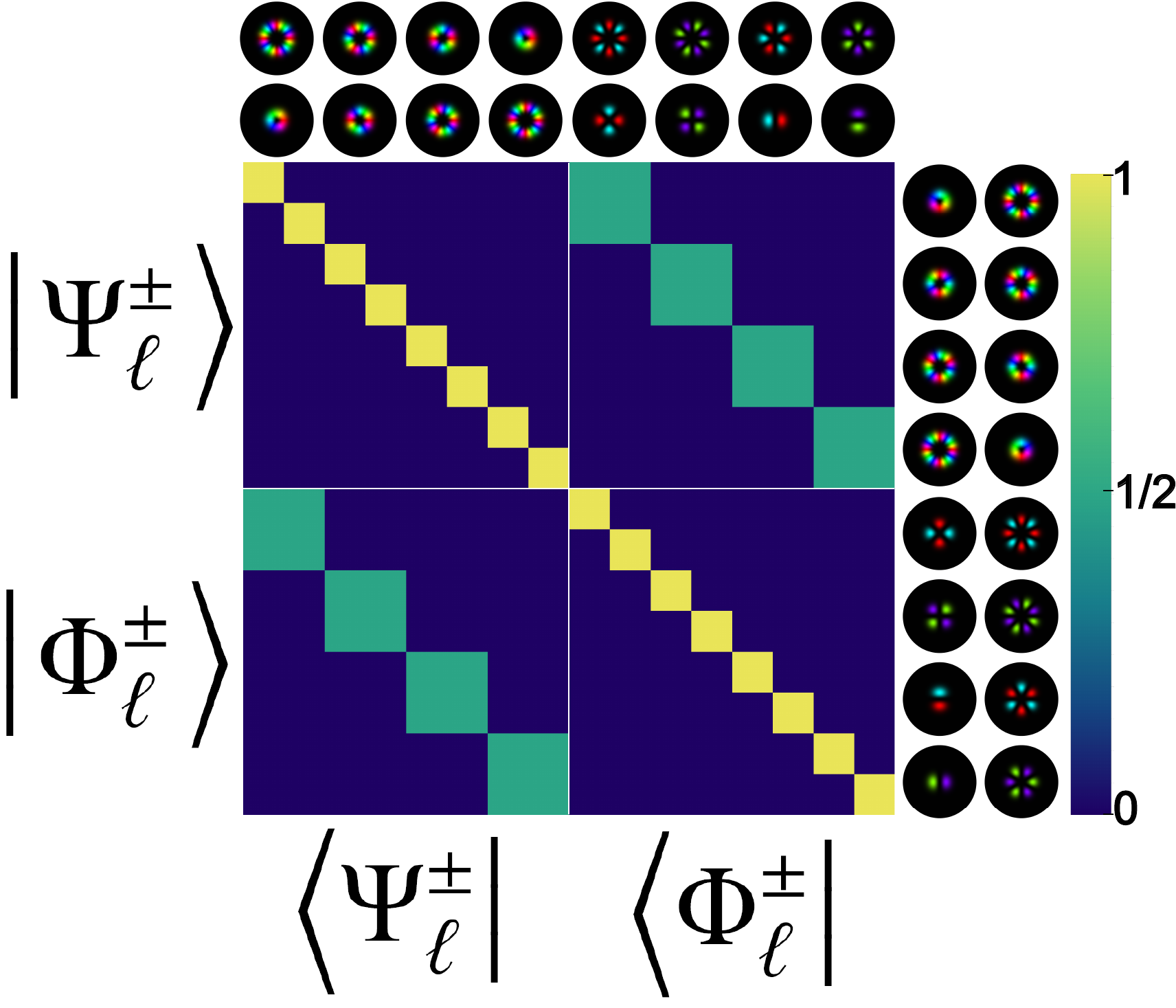}
		\caption{\textbf{Original modes in an 8-dimensional OAM-based protocol.} The originally conceptualized set of modes and the probability of detection matrices.  False color is given to the optical modes to represent the relative phase information.} 
		\label{fig:Supp:OLDPubs}
\end{figure}

These modes were initially conceptualized to behave well with an adaptive optics (AO) system. Previous work using our AO system has indicated that it will have a larger impact correcting modes with cylindrical symmetry and uniform intensity distributions~\cite{Scarfe2025}. Looking to make a QKD protocol that takes advantage of these properties, we designed a d-dimensional orthogonal basis of modes with cylindrical symmetry and uniform intensity distributions, see Fig.~\ref{fig:Supp:OLDPubs}. These modes are a subset of the full \Fqb{} basis where $j=-k=\ell$, and $m\in\{0,d/2-1\}$, given in Eq.~\ref{eq:OldPubs}  

\begin{equation}\label{eq:OldPubs}
    \ket{\phi_{\ell}^{\pm}}=\frac{1}{\sqrt{2}}\big(\ket{\ell} \pm \ket{-\ell}\big)
\end{equation}

The error rates in the noisy lab-scale turbulent environment shows that these modes and those in the logical basis are both affected by turbulence and corrected by the AO system equally. The level of ``Unbiasedness" of these two bases, however, is too low to reasonably perform any QKD experiment guaranteeing security. The idea of a encoding high dimensional information in a ``qubit-like" state stayed and the more general set of \Fqb{} modes discussed in the paper, $\ket{\phi_{j,k}^m}$, were conceived.

\subsection{Normalization of \Fqb{} Measurements}

For the normalization of the probability outcomes from measurements in the qubit-like basis, we use the completeness relation of the $|\phi_{jk}^{m}\rangle$ states:

\begin{eqnarray}
   && \sum_m \sum_{j<k} |\phi_{jk}^{(m)}\rangle \langle \phi_{jk}^{(m)} | = \frac{1}{2} \sum_m \sum_{j<k} \left( |j\rangle + \omega_d^m |k\rangle \right) \left( \langle j|+ \omega_d^{-m} \langle k | \right) \nonumber \\ \nonumber
    &=& \frac{1}{2} \sum_m \sum_{j<k} \left( |j\rangle \langle j| + \omega_d^{-m}|j\rangle \langle k| + \omega_d^m |k \rangle \langle j| +|k\rangle \langle k| \right) \\ \nonumber
    &=& \frac{1}{2} \sum_m \sum_{j<k} \left( |j\rangle \langle j| + |k\rangle \langle k|  \right) \\ \nonumber
    &=& \frac{1}{2} d \sum_{j<k} \left( |j\rangle \langle j| + |k\rangle \langle k|  \right) \\ \nonumber
    &=& \frac{1}{2} d (d-1) \hat{I},
\end{eqnarray}
where $\hat{I}$ is the $d$-dimensional identity matrix and where we have used the property that $\sum_m \omega_d^m =0$ and $\sum_m \omega_d^{-m} =0$. This completeness relation can then be used to normalized raw counts.

\begin{eqnarray}
    \sum_{m'} \sum_{j'<k'} p(j',k',m'|\hat{\rho}) &=&  \sum_{m'} \sum_{j'<k'} \mathrm{Tr} \left[ |\phi_{j'k'}^{(m')} \rangle \langle \phi_{j'k'}^{(m')} | \cdot \hat{\rho} \right] \nonumber \\ \nonumber
    &=&   \mathrm{Tr} \left[ \sum_m \sum_{j<k} |\phi_{j'k'}^{(m')} \rangle \langle \phi_{j'k'}^{(m')} | \cdot \hat{\rho} \right] \\ \nonumber
    &=&   \mathrm{Tr} \left[ \frac{1}{2} d (d-1) \hat{I} \cdot \hat{\rho} \right] \\ \nonumber
    &=& \frac{1}{2} d (d-1) \mathrm{Tr} \left[ \hat{\rho} \right] \\
    &=& \frac{1}{2} d (d-1).
\end{eqnarray}

Finally, the normalization condition is explicitly given by

\begin{eqnarray}
     \sum_{m'} \sum_{j'<k'}  p(j',k',m'|\hat{\rho}) = \frac{d(d-1)}{2}. \nonumber \\
\end{eqnarray}

\subsection{Derivation of inversion relation of Equation (8)}\label{Supp:Inver}

We start from the following relations linking the Fourier basis to the qubit-like basis,

\begin{eqnarray} \label{supp:eq:coeffs1}
    && p(\phi_{j'k'}^{m'}|\phi_{jk}^{m}) \\ \nonumber
     \nonumber && = \frac{4}{d^2} \sum_{\ell n} \cos^2 \left[\frac{\pi(m-(k-j)\ell)}{d} \right] \\ && \nonumber \ \ \ \ \ \ \ \ \ \ \ \ \ \ \  \cos^2 \left[\frac{\pi(m'-(k'-j')n)}{d} \right]  p(\eta_n|\eta_\ell).\\ \nonumber
\end{eqnarray}

This relation can also be inverted by finding the coefficients $\alpha_{jj'kl'mm'}^{(\ell,n)}$ such that

\begin{eqnarray} \label{supp:eq:alphacoeffs}
    p(\eta_n|\eta_\ell) = \sum_{j<k}\sum_{j'<k'}\sum_{m,m'} \alpha_{jj'kk'mm'}^{(\ell,n)} p(\phi_{j'k'}^{m'}|\phi_{jk}^{m}). \nonumber
\end{eqnarray} 

We now derive an analytical expression for $\alpha_{jj'kk'mm'}^{(\ell,n)}$ by starting with the assumption that it can be written as the following product with appropriate separation of variables, i.e.,

\begin{eqnarray}
    \alpha_{jj'kk'mm'}^{(\ell,n)} = \alpha_{jkm}^{(\ell)} \alpha_{j'k'm'}^{(n)}.
\end{eqnarray}

Let us also define the following expressions,

\begin{eqnarray}
    \beta_{jkm}^{(\ell)} = \frac{2}{d}\cos^2 \left[\frac{\pi(m-(k-j)\ell)}{d} \right].
\end{eqnarray}

By doing so, we get the following forward relation,

\begin{eqnarray} \label{supp:eq:coeffs2}
     p(\phi_{j'k'}^{m'}|\phi_{jk}^{m})  = \sum_{\ell n} \beta_{jkm}^{(\ell)} \beta_{j'k'm'}^{(n)}  p(\eta_n|\eta_\ell).\\ \nonumber
\end{eqnarray}

By multiplying both side of this equation by $\alpha_{jkm}^{(\ell')} \alpha_{j'k'm'}^{(n')}$ and summing over $j<k$, $j'<k'$, $m$ and $m'$, we get,

\begin{eqnarray} \label{supp:eq:coeffs3}
     \sum_{j<k} \sum_{j'<k'} \sum_{m,m'} \alpha_{jkm}^{(\ell')} \alpha_{j'k'm'}^{(n')} p(\phi_{j'k'}^{m'}|\phi_{jk}^{m})  = \nonumber \\ \nonumber \sum_{j<k} \sum_{j'<k'} \sum_{m,m'} \sum_{\ell n} \alpha_{jkm}^{(\ell')} \alpha_{j'k'm'}^{(n')} \beta_{jkm}^{(\ell)} \beta_{j'k'm'}^{(n)}  p(\eta_n|\eta_\ell).\\ \nonumber
\end{eqnarray}

We thus need to find $\alpha_{jkm}^{(\ell')}$ and $\alpha_{jkm'}^{(n')}$, such that,

\begin{eqnarray}
    \sum_{j<k}\sum_m \alpha_{j,k,m}^{(\ell')} \beta_{j,k,m}^{(\ell)} = \delta_{\ell,\ell'} \nonumber \\
    \sum_{j'<k'}\sum_{m'} \alpha_{j',k',m'}^{(n')} \beta_{j',k',m'}^{(n)} = \delta_{n,n'}, \nonumber
\end{eqnarray}

to retrieve the backward relation, i.e.,

\begin{eqnarray} \label{supp:eq:coeffs4}
     \sum_{j<k} \sum_{j'<k'} \sum_{m,m'} \alpha_{jkm}^{(\ell')} \alpha_{j'k'm'}^{(n')} p(\phi_{j'k'}^{m'}|\phi_{jk}^{m})  = \nonumber \\  \sum_{\ell n} \delta_{\ell,\ell'} \delta_{n,n'}  p(\eta_n|\eta_\ell) = p(\eta_{n'}|\eta_{\ell'}). \nonumber \\
\end{eqnarray}

Let us try the following solution for $\alpha_{j,k,m}^{(\ell)}$,

\begin{eqnarray}
    \alpha_{j,k,m}^{(\ell)} = \frac{2}{d^2} \left( \frac{2-d}{d-1} + \sum_{p=0}^{d-1} \exp \left( \frac{2 \pi i}{d} p \left( m - (k-j)\ell \right) \right) \right).
\end{eqnarray}

We now need to show that,

\begin{eqnarray}
    \sum_{j<k}\sum_m \alpha_{j,k,m}^{(\ell')} \beta_{j,k,m}^{(\ell)} = \delta_{\ell,\ell'}. \nonumber 
\end{eqnarray}

Inserting the explicit form of $\alpha_{j,k,m}^{(\ell')}$ and $\beta{j,k,m}^{(\ell)}$, we get

\begin{eqnarray}
    && \sum_{j<k}\sum_m \alpha_{j,k,m}^{(\ell')} \beta_{j,k,m}^{(\ell)} \nonumber \\
    &=& \sum_{j<k} \sum_m \left( \frac{2}{d^2} \left( \frac{2-d}{d-1} + \sum_{p=0}^{d-1} \exp \left( \frac{2 \pi i}{d} p \left( m - (k-j)\ell' \right) \right) \right) \right) \nonumber \\
   && \ \ \ \ \  \left( \frac{2}{d}\cos^2 \left(\frac{\pi(m-(k-j)\ell)}{d} \right) \right) \nonumber \\
   &=& \frac{2}{d^3}  \sum_{j<k} \sum_m \left( \frac{2-d}{d-1} + \sum_{p=0}^{d-1} \exp \left( \frac{2 \pi i}{d} p \left( m - (k-j)\ell' \right) \right) \right) \nonumber \\
   && \ \ \ \ \ \left(1 + \cos \left(\frac{2 \pi }{d} \left( m - (k-j)\ell \right)  \right) \right) \nonumber \\
   &=& \frac{2}{d^3}  \sum_{j<k} \sum_m \left( \frac{2-d}{d-1} + \sum_{p=0}^{d-1} \exp \left( \frac{2 \pi i}{d} p \left( m - (k-j)\ell' \right) \right) \right) \nonumber \\
   && \ \ \left(1 + \frac{1}{2} \exp \left(\frac{2 \pi i }{d} \left( m - (k-j)\ell \right)  \right) +  \frac{1}{2} \exp \left(\frac{-2 \pi i }{d} \left( m - (k-j)\ell \right)  \right) \right) \nonumber. \\ 
   &=& \frac{2}{d^3}  \sum_{j<k} \Bigg[ \sum_m \left( \frac{2-d}{d-1} \right) + \frac{1}{2} \left( \frac{2-d}{d-1} \right) \sum_m \exp \left(\frac{2 \pi i }{d} \left( m - (k-j)\ell \right)  \right) \nonumber \\
   && + \frac{1}{2} \left( \frac{2-d}{d-1} \right) \sum_m \exp \left(\frac{-2 \pi i }{d} \left( m - (k-j)\ell \right)  \right) \nonumber \\
   && + \sum_m \sum_p  \exp \left( \frac{2 \pi i}{d} p \left( m - (k-j)\ell' \right) \right) \nonumber \\
   && + \frac{1}{2} \sum_m \sum_p  \exp \left( \frac{2 \pi i}{d} p \left( m - (k-j)\ell' \right) \right) \exp \left(\frac{2 \pi i }{d} \left( m - (k-j)\ell \right)  \right) \nonumber \\
   && + \frac{1}{2} \sum_m \sum_p  \exp \left( \frac{2 \pi i}{d} p \left( m - (k-j)\ell' \right) \right) \exp \left(\frac{-2 \pi i }{d} \left( m - (k-j)\ell \right)  \right) \Bigg] \nonumber \\
\end{eqnarray}

We can now make use of the following orthogonality relation,

\begin{eqnarray}
    \sum_{m=0}^{d-1} \exp \left( \frac{2\pi i}{d} m s \right) = d \delta_{s,0}. \nonumber
\end{eqnarray}

Continuing with our derivation, 

\begin{eqnarray}
    && \sum_{j<k}\sum_m \alpha_{j,k,m}^{(\ell')} \beta_{j,k,m}^{(\ell)} \nonumber \\
   &=& \frac{2}{d^3}  \sum_{j<k} \Bigg[ d \left( \frac{2-d}{d-1} \right)  \nonumber \\
   && + \sum_p d \delta_{p,0} \exp \left( \frac{-2 \pi i}{d} p (k-j)\ell'  \right) \nonumber \\
   && + \frac{1}{2}  \sum_p d \delta_{p,d-1} \exp \left( \frac{-2 \pi i}{d} p  (k-j)\ell'  \right) \exp \left(\frac{-2 \pi i }{d} (k-j)\ell   \right) \nonumber \\
   && + \frac{1}{2} \sum_p d \delta_{p,1} \exp \left( \frac{-2 \pi i}{d} p  (k-j)\ell' \right) \exp \left(\frac{2 \pi i }{d}  (k-j)\ell   \right) \Bigg] \nonumber \\
   &=& \frac{2}{d^2}  \sum_{j<k} \Bigg[  \left( \frac{2-d}{d-1} \right) +1   \nonumber \\
   && + \frac{1}{2}     \exp \left( \frac{2 \pi i}{d}   (k-j)(\ell'-\ell)  \right)  + \frac{1}{2} \exp \left( \frac{-2 \pi i}{d} (k-j)(\ell' - \ell) \right) \Bigg]\nonumber \\
   &=& \frac{2}{d^2}  \sum_{j<k} \Bigg[  \left( \frac{1}{d-1} \right)  + \cos \left( \frac{2 \pi}{d}  (k-j)(\ell'-\ell)  \right)  \Bigg].\nonumber \\
\end{eqnarray}

We note here that the expression in the sum only depends on the difference between $j$ and $k$, namely $r=k-j$. We can then re-arrange the sum over $j$ and $k$ as follows,

\begin{eqnarray}
    \sum_{j<k} \longrightarrow \sum_{r=1}^{d-1} (d-r). \nonumber
\end{eqnarray}

We now have,

\begin{eqnarray}
     && \sum_{j<k}\sum_m \alpha_{j,k,m}^{(\ell')} \beta_{j,k,m}^{(\ell)} \nonumber \\ 
   &=& \frac{2}{d^2}  \sum_{r=1}^{d-1} (d-r) \Bigg[  \left( \frac{1}{d-1} \right)  + \cos \left( \frac{2 \pi}{d} r (\ell'-\ell)  \right)  \Bigg].\nonumber \\
\end{eqnarray}

In order to show that the right-hand side here is equal to $\delta_{\ell,\ell'}$, let us consider two distinct cases, i.e., the case where $\ell=\ell'$ and $\ell \neq \ell'$. If we find that the right hand side is equal to 1 in the former case and 0 in the latter case, we have thus proven that it is equal to the Kronecker delta function, $\delta_{\ell,\ell'}$.

We start with the first case of $\ell=\ell'$.

\begin{eqnarray}
     && \sum_{j<k}\sum_m \alpha_{j,k,m}^{(\ell)} \beta_{j,k,m}^{(\ell)} \nonumber \\ 
   &=& \frac{2}{d^2}  \sum_{r=1}^{d-1} (d-r) \Bigg[  \left( \frac{1}{d-1} \right)  + 1  \Bigg]\nonumber \\
   &=&   \frac{2}{d^2} \left( \frac{d}{d-1} \right)  \sum_{r=1}^{d-1} (d-r) \nonumber \\
   &=&   \frac{2}{d} \left( \frac{1}{d-1} \right) \left( d (d-1) - \frac{(d-1)d}{2} \right) \nonumber \\
   &=& 1. 
\end{eqnarray}

We now need to show the second case where $\ell \neq \ell'$.

\begin{eqnarray}
     && \sum_{j<k}\sum_m \alpha_{j,k,m}^{(\ell')} \beta_{j,k,m}^{(\ell)} \nonumber \\ 
   &=& \frac{2}{d^2}   \sum_{r=1}^{d-1} (d-r) \Bigg[  \left( \frac{1}{d-1} \right)  + \cos \left( \frac{2 \pi}{d} r (\ell'-\ell)  \right)  \Bigg] \nonumber \\
   &=& \frac{2}{d^2} \Bigg[ \frac{1}{d-1} \sum_{r=1}^{d-1} (d-r) \nonumber \\
   && \ \ \ + \sum_{r=0}^{d-1} (d-r) \cos \left( \frac{2\pi}{d} r (\ell'-\ell) \right) - d \nonumber \Bigg] \\
   &=& \frac{2}{d^2} \Bigg[ \frac{1}{d-1} \frac{d(d-1)}{2} + \sum_{r=0}^{d-1} (d-r) \cos \left( \frac{2\pi}{d} r (\ell'-\ell) \right) - d \nonumber \Bigg] \\
   &=& \frac{2}{d^2} \Bigg[- \frac{d}{2}  + \sum_{r=0}^{d-1} (d-r) \cos \left( \frac{2\pi}{d} r (\ell'-\ell) \right)  \nonumber \Bigg] \nonumber \\
    &=& \frac{1}{d^2} \Bigg[- d  + \sum_{r=0}^{d-1} (d-r) \exp \left( \frac{2\pi i}{d} r (\ell'-\ell) \right) \nonumber \\
    && \ \ \ \ \ \ \ \  \ \ + \sum_{r=0}^{d-1} (d-r) \exp \left( \frac{-2\pi i}{d} r (\ell'-\ell) \right)  \nonumber \Bigg]. \nonumber \\
\end{eqnarray}

If we take a closer look now at the following sum, i.e.,

\begin{eqnarray}
    \sum_{r=0}^{d-1} (d-r) \exp \left( \frac{2\pi i}{d} r (\ell'-\ell) \right),
\end{eqnarray}

We realize that is not quite the orthogonality relation we are used to. We can try to calculate the sum in our case of $\ell \neq \ell'$.

We start by defining a generating function, $G(z)$, given by,

\begin{eqnarray}
    G(z) = \sum_{r=0}^{d-1} (d-r) z^r \nonumber \\
    = \sum_{r=0}^n (n+1-r) z^r, 
\end{eqnarray}

where we have defined $n=d-1$. By doing so, we can use the well known result for finite geometric series, namely,

\begin{eqnarray}
    S(z) = \sum_{r=0}^n z^r = \frac{1-z^{n+1}}{1-z},
\end{eqnarray}

which is true for $z\neq 1$, which also holds true since $\ell \neq \ell'$. 

Our generating function then takes the following form,

\begin{eqnarray}
    G(z) = \sum_{r=0}^n (n+1-r) z^r = (n+1) S(z) - \sum_{r=0}^n r\  z^r.
\end{eqnarray}

We can now differentiate $S(z)$ with respect to $z$, 

\begin{eqnarray}
    S'(z) = \sum_{r=0}^n r z^{r-1}. 
\end{eqnarray}

Thus, we can rewrite our generating function as, 

\begin{eqnarray}
    G(z) = (n+1) S(z) - z S'(z).
\end{eqnarray}

Let us now perform the differentiation of $S(z)$ explicitly,

\begin{eqnarray}
    S'(z) &=& \frac{-(n+1) z^n}{1-z}+ \frac{1-z^{n+1}}{(1-z)^2} \nonumber \\
    &=& \frac{-(n+1) z^n (1-z) + (1-z^{n+1})}{(1-z)^2}. 
\end{eqnarray}

Putting everything together, we get,

\begin{eqnarray}
    G(z) &=& \frac{(n+1) (1-z^{n+1})}{1-z} + \frac{(n+1) z^{n+1} (1-z) - z (1-z^{n+1})}{(1-z)^2} \nonumber \\
    &=& \frac{(n+1)(1-z^{n+1})(1-z) + (n+1) z^{n+1} (1-z) -z (1-z^{n+1})}{(1-z)^2} \nonumber \\
    &=& \frac{(n+1)(1-z) - z(1-z^{n+1})}{(1-z)^2}.
\end{eqnarray}

Going back to $d$, we have, 

\begin{eqnarray}
    G(z) = \frac{d(1-z)-z(1-z^d)}{(1-z)^2}.
\end{eqnarray}

At this point, we remind ourselves that $z=\exp(2 \pi i (\ell-\ell')/d )$. Thus, we can see that,

\begin{eqnarray}
    z^d=1.
\end{eqnarray}

Thus, our generating function greatly simplifies to,

\begin{eqnarray}
    G(z) = \frac{d}{1-z}. 
\end{eqnarray}

Going back to our equation,

\begin{eqnarray}
     && \sum_{j<k}\sum_m \alpha_{j,k,m}^{(\ell')} \beta_{j,k,m}^{(\ell)} \nonumber \\ 
    &=& \frac{1}{d^2} \Bigg[- d  + \sum_{r=0}^{d-1} (d-r) \exp \left( \frac{2\pi i}{d} r (\ell'-\ell) \right) \nonumber \\
    && \ \ \ \ \ \ \ \  \ \ + \sum_{r=0}^{d-1} (d-r) \exp \left( \frac{-2\pi i}{d} r (\ell'-\ell) \right)  \nonumber \Bigg]. \nonumber \\
    &=& \frac{1}{d^2} \left[ -d +\sum_{r=0}^{d-1}(d-r)z^r +\sum_{r=0}^{d-1}(d-r)(z^*)^r  \right] \nonumber \\
    &=& \frac{1}{d^2} \left[ -d +\frac{d}{1-\exp \left( \frac{2\pi i}{d} r (\ell'-\ell) \right)} +\frac{d}{1-\exp \left( \frac{-2\pi i}{d} r (\ell'-\ell) \right)}   \right] \nonumber \\
    &=& \frac{1}{d} \left[ -1 +\frac{\left( 1-\exp \left( \frac{-2\pi i}{d} r (\ell'-\ell) \right)\right) +\left( 1-\exp \left( \frac{-2\pi i}{d} r (\ell'-\ell) \right) \right)}{\left(1-\exp \left( \frac{2\pi i}{d} r (\ell'-\ell) \right) \right) \left( 1-\exp \left( \frac{-2\pi i}{d} r (\ell'-\ell) \right) \right)}  \right] \nonumber \\
    &=& \frac{1}{d} \left[ -1 +\frac{1-\cos \left( \frac{2 \pi }{d} (\ell - \ell') \right)}{1-\cos \left( \frac{2 \pi }{d} (\ell - \ell') \right)}  \right] \nonumber \\
    &=& \frac{1}{d} \left[ -1 +1  \right] \nonumber \\
    &=& 0. \nonumber
\end{eqnarray}

We have thus proven that,

\begin{eqnarray}
     \sum_{j<k}\sum_m \alpha_{j,k,m}^{(\ell')} \beta_{j,k,m}^{(\ell)} \ = \delta_{\ell,\ell'},
\end{eqnarray}

for all $\ell$ and $\ell'$.

Our final forward and backward relations linking the Fourier basis to the qubit-like basis is given by the following,

\begin{eqnarray} \label{supp:eq:coeffs5}
    && p(\phi_{j'k'}^{m'}|\phi_{jk}^{m}) \\ \nonumber
     \nonumber && = \frac{4}{d^2} \sum_{\ell n} \cos^2 \left[\frac{\pi(m-(k-j)\ell)}{d} \right] \\ && \nonumber \ \ \ \ \ \ \ \ \ \ \ \ \ \ \  \cos^2 \left[\frac{\pi(m'-(k'-j')n)}{d} \right]  p(\eta_n|\eta_\ell).\\ \nonumber
\end{eqnarray}

\begin{eqnarray} \label{supp:eq:coeffs6}
   && p(\eta_n|\eta_\ell) = \frac{4}{d^4} \sum_{j<k}\sum_{j'<k'}\sum_{m,m'} \left( \frac{2-d}{d-1} + \sum_{p=0}^{d-1} \exp \left( \frac{2 \pi i}{d} p \left( m - (k-j)\ell \right) \right) \right) \nonumber \\
    && \left( \frac{2-d}{d-1} + \sum_{p'=0}^{d-1} \exp \left( \frac{2 \pi i}{d} p' \left( m' - (k'-j')n \right) \right) \right) p(\phi_{j'k'}^{m'}|\phi_{jk}^{m}).
\end{eqnarray}

\subsection{Full Proof of Security}

In the case of an ideal single photon source, we consider the action of an eavesdropper, Eve, as a general collective attack in the Fourier basis given by the unitary transformation $U_\mathrm{Eve}$, i.e.,

\begin{eqnarray}
    U_\mathrm{Eve} |\eta_i\rangle |e_{00}\rangle = \sum_{j=0}^{d-1} c_{i j} |\eta_j\rangle |e_{i j}\rangle,
\end{eqnarray}

where $|e_{i j}\rangle$ is Eve's ancilla state. Without loss of generality, we assume that $c_{ij} \geq 0$ and $\langle e_{ij}|e_{mn}\rangle=\delta_{im}\delta_{jn}$. To estimate Eve's leaked information for the case where Alice encodes and Bob measures states, $|\psi_i\rangle$ in the computational basis, we can consider the error rate in the hypothetical case where Alice encodes and Bob measures states, ${|\eta_i \rangle=\frac{1}{\sqrt{d}}\sum_{n=0}^{d-1} \omega_d^{i n} |n\rangle }$, from a mutually unbiased basis (MUB), which we call here the Fourier basis. This hypothetical dit error rate, $E_d'$, depends on the probability that Bob obtains a measurement of $|\eta_{i'} \rangle$ conditioned on the fact that Alice prepares her state as $|\eta_i \rangle$, i.e.,

\begin{eqnarray}
   && p(i' | i) \\
   \nonumber && = \mathrm{Tr} \left[ |\eta_{i'}\rangle \langle \eta_{i'}| U_\mathrm{Eve} |\eta_{i}\rangle \langle \eta_{i}| \otimes |e_{00}\rangle \langle e_{00}| U_\mathrm{Eve}^\dagger \right] \\
   \nonumber && = \sum_{j,j'} \mathrm{Tr} \left[ |\eta_{i'} \rangle \langle \eta_{i'} | c_{i j} |\eta_j \rangle |e_{i j} \rangle \langle \eta_{j'} | \langle e_{i j'}| c_{i j'} \right] \\
   \nonumber && = \mathrm{Tr} \left[ c_{i i'}^2 |e_{i i'} \rangle \langle e_{i i'} | \right] \\
   \nonumber && = c_{i i'}^2
\end{eqnarray}

We are now interested in relating the probability outcomes $p(i'|i)$ in the Fourier basis to the measured probability outcomes of the qubit-like states. To do so, we use the fact that the qubit-like states can be rewritten in terms of the Fourier basis states,

\begin{eqnarray}
    |\phi_{jk}^{(m)}\rangle = \frac{1}{\sqrt{2d}}\sum_\ell \left( \omega_d^{-j \ell}+\omega_d^{(m-k \ell)} \right) |\eta_\ell \rangle.
\end{eqnarray}

The probability that Bob obtains $|\phi_{j'k'}^{(m')}\rangle$ conditioned on the fact that Alice prepares her state as $|\phi_{ij}^{(m)}\rangle$ is given by,

\begin{eqnarray}
    && p(\phi_{j'k'}^{m'}|\phi_{jk}^{m}) \\ 
    \nonumber && = \mathrm{Tr} \left[ |\phi_{j'k'}^{(m')}\rangle \langle \phi_{j'k'}^{(m')}| U_\mathrm{Eve} |\phi_{jk}^{(m)}\rangle \langle \phi_{jk}^{(m)}| \otimes |e_{00}\rangle \langle e_{00}| U_\mathrm{Eve}^\dagger \right] \\
    \nonumber && = \frac{1}{2d} \sum_{\ell \ell'} \left( \omega_d^{-j\ell} + \omega_d^{(m-k\ell)} \right) \left( \omega_d^{j \ell'}+\omega_d^{(k\ell'-m)} \right) \\ && \nonumber \ \ \ \ \ \ \ \ \mathrm{Tr} \left[  |\phi_{j'k'}^{(m')}\rangle \langle \phi_{j'k'}^{(m')}| U_\mathrm{Eve} |\eta_\ell\rangle \langle \eta_{\ell'} | \otimes |e_{00}\rangle \langle e_{00}| U_\mathrm{Eve}^\dagger  \right]  \\
    \nonumber && = \frac{1}{2d} \sum_{\ell \ell'} \sum_{n n'} \left( \omega_d^{-j \ell} + \omega_d^{(m-k\ell)} \right) \left( \omega_d^{j \ell'}+\omega_d^{(k\ell'-m)} \right) \\ && \nonumber \ \ \ \ \ \ \ \ \mathrm{Tr} \left[  |\phi_{j'k'}^{(m')}\rangle \langle \phi_{j'k'}^{(m')}| c_{\ell n} |\eta_n\rangle |e_{\ell n} \rangle \langle \eta_{n'} \langle e_{\ell' n'}| c_{\ell'n'} \right]  \\
    \nonumber && = \left(\frac{1}{2d}\right)^2 \sum_{\ell \ell'} \sum_{n n'} \sum_{ss'} \left( \omega_d^{-j\ell} + \omega_d^{(m-k\ell)} \right) \left( \omega_d^{j \ell'}+\omega_d^{(k\ell'-m)} \right) \\ && \nonumber\ \ \ \ \ \  \left( \omega_d^{-j's} + \omega_d^{(m'-k's)} \right) \left( \omega_d^{j' s'}+\omega_d^{(k's'-m')} \right) \\ && \nonumber \ \ \ \ \ \ \ \ \ \ \mathrm{Tr} \left[  |\eta_{s'}\rangle \langle \eta_s| c_{\ell n} |\eta_n\rangle |e_{\ell n} \rangle \langle \eta_{n'} \langle e_{\ell' n'}| c_{\ell'n'} \right]  \\
    \nonumber && = \left(\frac{1}{2d}\right)^2 \sum_{\ell \ell'} \sum_{n n'} \left( \omega_d^{-j\ell} + \omega_d^{(m-k\ell)} \right) \left( \omega_d^{j \ell'}+\omega_d^{(k\ell'-m)} \right) \\ && \nonumber\ \ \ \ \ \  \left( \omega_d^{-j'n} + \omega_d^{(m'-k'n)} \right) \left( \omega_d^{j' n'}+\omega_d^{(k'n'-m')} \right) \\ && \nonumber \ \ \ \ \ \ \ \ \ \ \mathrm{Tr} \left[   c_{\ell n} c_{\ell'n'} |e_{\ell n} \rangle \langle e_{\ell' n'}|  \right]  \\
    \nonumber && = \left(\frac{1}{2d}\right)^2 \sum_{\ell n} \left( \omega_d^{-j\ell} + \omega_d^{(m-k\ell)} \right) \left( \omega_d^{j \ell}+\omega_d^{(k\ell-m)} \right) \\ && \nonumber\ \ \ \ \ \  \left( \omega_d^{-j'n} + \omega_d^{(m'-k'n)} \right) \left( \omega_d^{j' n}+\omega_d^{(k'n-m')} \right) c_{\ell n}^2
\end{eqnarray}

Simplyfing,

\begin{eqnarray}
    && p(\phi_{j'k'}^{m'}|\phi_{jk}^{m}) \\ 
    \nonumber && = \left(\frac{1}{2d}\right)^2 \sum_{\ell n} \left(1 + \omega_d^{(m-k\ell+j\ell)} \right) \left( 1+ \omega_d^{-(m-k\ell+j\ell)} \right) \\ && \nonumber\ \ \ \ \ \  \left(1 + \omega_d^{(m'-k'n+j'n)} \right) \left( 1+ \omega_d^{-(m'-k'n+j'n)} \right)  c_{\ell n}^2 \\
    \nonumber && = \left(\frac{1}{d}\right)^2 \sum_{\ell n} \left(1 + \cos \left(\frac{2\pi(m-k\ell+j\ell)}{d} \right) \right) \\ && \nonumber \ \ \ \ \ \  \left(1 + \cos \left(\frac{2\pi(m'-k'n+j'n)}{d} \right) \right)  c_{\ell n}^2
\end{eqnarray}

We have then arrived to a relation between the outcome probabilities in the Fourier basis, $p(n|\ell))$, necessary to determine the dit error rate $E_d'$, and the qubit-like probability outcomes, $p(\phi_{j'k'}^{m'}|\phi_{jk}^{m})$, measured experimentally, i.e.,

\begin{eqnarray}
    && p(\phi_{j'k'}^{m'}|\phi_{jk}^{m}) \\ \nonumber
     \nonumber && = \frac{4}{d^2} \sum_{\ell n} \cos^2 \left[\frac{\pi(m-(k-j)\ell)}{d} \right] \\ && \nonumber \ \ \ \ \ \ \ \ \ \ \ \ \ \ \  \cos^2 \left[\frac{\pi(m'-(k'-j')n)}{d} \right]  p(n|\ell) \\ \nonumber
\end{eqnarray}

This can be inverted by finding the coefficients $\alpha_{jj'kk'mm'}^{(\ell,n)}$ such that,

\begin{eqnarray}
    p(n|\ell) = \sum_{j<k}\sum_{j'<k'}\sum_{m,m'} \alpha_{jj'kk'mm'}^{(\ell,n)} p(\phi_{j'k'}^{m'}|\phi_{jk}^{m})
\end{eqnarray}

The exact values of $\alpha_{jj'kk'mm'}^{(\ell,n)}$ are given in Equation \ref{supp:eq:coeffs6}. We note that these are the computational and Fourier bases form the two bases required for the high-dimensional BB84 protocol. Thus, Eve's leaked information on Alice encoding and Bob measuring in the computational basis is given by $I_\mathrm{AE}\leq h^{(d)}(E'_d)$, where we have used the fact that the phase error in the computational basis is equal to the dit error rate in the Fourier basis. Explicitly, the dit error rate in the Fourier basis is given by,

\begin{eqnarray}
    E_d' &=& \frac{1}{d} \sum_{\ell \neq n} p(n|\ell) \\ \nonumber
    &=& \frac{1}{d} \left( \sum_{\ell n} p(n|\ell) - \sum_\ell p(\ell|\ell)  \right)
\end{eqnarray}

Finally, the secret key rate per sifted key for Alice and Bob is given by,

\begin{eqnarray}
    R = \log_2(d) - h^{(d)} (E_d) - I_\mathrm{AE} \geq \log_2(d) - h^{(d)} (E_d) - h^{(d)} (E'_d).
\end{eqnarray}

\end{document}